\documentclass[english]{article}

\usepackage{geometry}
\usepackage{graphicx}
\usepackage{amssymb}
\usepackage{epstopdf}
\usepackage{babel}
\usepackage{inputenc}
\usepackage{epsfig}

\begin{document}

\begin{center}
{\Large {\bf Atmospheric  $\nu$ and Long Baseline $\nu$ experiments II}} 
\vspace{0.8cm}

\par~\par

{\author G{G}. Giacomelli \\
Dept. of Physics, Univ. of Bologna , and INFN\\
 V.le C. Berti Pichat 6/2, Bologna, I-40127, Italy\\
 E-mail: giacomelli@bo.infn.it }\\

\par~\par

Invited Lecture at the Carpatian Summer School of Physics 2010, Sinaia, Romania, 20 June - 3 July 2010.

\end{center}

\vspace{0.5cm}

 {\bf Abstract.} {A short introduction summarizes the experimental high energy field of neutrino oscillations and the oscillation notations. Then the results obtained by experiments on atmospheric neutrino oscillations are summarized and discussed. At the beginning of the new century a number of long baseline neutrino beams became available. The  results obtained by different long baseline neutrino oscillation experiments are considered, including the first possible event $\nu_{\mu} \rightarrow \nu_{\tau}$ in appearance mode. Finally conclusions and perspectives are discussed.} 
 
\vspace{0.5cm}

 
\section{Introduction}
A high energy primary cosmic ray (CR), proton or nucleus, interacts in the upper atmosphere producing a large number of charged pions and kaons, which decay yielding muons and $\nu_{\mu}$'s; also the muons decay yielding neutrinos $\nu_{\mu}$ and $\nu_{e}$. The ratio of the numbers of $\nu_{\mu}$ to $\nu_{e}$ is $\simeq$2 and N$_{\nu}$/N$_{\overline{\nu}}$ $\simeq$1. These atmospheric neutrinos are produced at 10-20 km above ground, and they proceed towards the earth.

Atmospheric neutrinos are suited for the study of neutrino oscillations, since they have energies from a fraction of GeV up to more than 100 GeV and they travel distances $L$ from few tens of km up to 13000 km; thus $L/E_{\nu}$ ranges from $\sim$1 km/GeV to $\sim$10$^{5}$ km/GeV.

The early water Cherenkov detectors IMB \cite{1} and Kamiokande \cite{2} reported anomalies in the ratio of muon to electron neutrinos, while the tracking calorimeters NUSEX \cite{3}, Frejus \cite{4}, and the Baksan \cite{5} scintillator detector did not find any. In 1995 MACRO found a deficit for upthrougoing muons \cite{6}. The Soudan 2 experiment \cite{7} confirmed the ratio anomaly. 

In 1998 Soudan 2 \cite{8}, MACRO \cite{9} and SuperKamiokande (SK) \cite{10} reported deficits in the $\nu_{\mu}$ fluxes with respect to Monte Carlo (MC) predictions and distortions in the angular distributions; instead the $\nu_{e}$ distributions were in agreement with non oscillated MCs. These features were explained in terms of $\nu_{\mu} \leftrightarrow \nu_{\tau}$ oscillations.

The MC atmospheric neutrino flux was theorically computed by many authors in the mid 1990s \cite{11} and in the early 2000s \cite{12}. The last computations had improvements, but also a new scale uncertainty.

Several long baseline $\nu$ beams became operational: KEK to Kamioka (250 km), NuMi from Fermilab to the Soudan mine (735 km) and CERN to Gran Sasso(GS) (CNGS) (730 km). The K2K and MINOS experiments obtained results in agreement with the atmospheric $\nu$ results. In the CNGS beam the OPERA experiment reported recently a possible first candidate event $\nu_{\mu} \leftrightarrow \nu_{\tau}$ in appearance mode.

Very recently the MINOS experiment reported a possible anomaly of $\overline{\nu_{\mu}}$ with respect to $\nu_{\mu}$ and on a possible indication for a 4$^{th}$ neutrino.
 
Long baseline experiments are expected to give increasing contributions to neutrino physics.

\section{Neutrino oscillations and masses}

If neutrinos have non-zero masses, one has to consider the 3 $weak$ $flavour$ $eigenstates$ $\nu_{e}$, $\nu_{\mu}$, $\nu_{\tau}$ and the 3 $mass$ $eigenstates$ $\nu_{1}$, $\nu_{2}$, $\nu_{3}$. The flavour eigenstates $\nu_{l}$ are linear combinations of the mass eigenstates $\nu_{m}$  (and viceversa) via the elements of the unitary mixing matrix $U_{lm}$:

\footnotesize
\begin{equation}
\nu_{l} = \sum_{m=1}^{3} U_{lm}\nu_{m}     
\end{equation}
\normalsize

In the conventional parametrization the matrix U reads as follows:
\footnotesize
\begin{equation}
U\equiv
        \left( \begin{array}{ccc}
        1 &      0  &     0  \\
        0 &  c_{23} & s_{23} \\
        0 & -s_{23} & c_{23}

\end{array} \right)
        \left( \begin{array}{ccc}
                     c_{13} & 0 &  s_{13}e^{i\delta}  \\
                          0 & 1 &                  0  \\
        -s_{13}e^{-i\delta} & 0 &             c_{13}
  \end{array} \right)
        \left( \begin{array}{ccc}
         c_{12} & s_{12} & 0 \\
        -s_{12} & c_{12} & 0 \\
       0  &      0 & 1
  \end{array} \right)
\end{equation}
\normalsize
with $s_{12}$ $\equiv$ $\sin{\theta_{12}}$, and similarly for the other sines and cosines. 

In the simple case of two flavour ($\nu_{\mu}$, $\nu_{\tau}$) and two mass eigenstates ($\nu_{2}$, $\nu_{3}$) one has

\begin{equation}
\left\{
\begin{array}{lr}
 \nu_{\mu} =  &  \nu_{2} \cos{\theta}_{23} + \nu_{3} \sin{\theta}_{23} \\
 \nu_{\tau} = & -\nu_{2} \sin{\theta}_{23} + \nu_{3} \cos{\theta}_{23}
\end{array}
\right.
\end{equation}
\normalsize 

{\noindent} where $\theta_{23}$ is the mixing angle.  The probability for an initial $\nu_{\mu}$ to oscillate into a $\nu_{\tau}$ is 

\begin{equation}
P(\nu_{\mu} \rightarrow \nu_{\tau}) = \sin^{2} 2\theta_{23} \sin^{2} \left( \frac{1.27 \Delta m_{23}^{2} L}{E_{\nu}} \right)
\end{equation}

The survival probability of a $\nu_{\mu}$ beam is

\begin{equation}
P(\nu_{\mu} \rightarrow \nu_{\mu}) 
\simeq 1 -  P(\nu_{\mu} \rightarrow \nu_{\tau}) =  1 - \sin^{2} 2 \theta_{23} \sin^{2} \left( \frac{1.27 \Delta m_{23}^{2} L}{E_{\nu}} \right)
\end{equation}
\normalsize

{\noindent} where $ \Delta m_{23}^{2}$ = $m_{3}^{2} - m_{2}^{2}$ and L is the distance travelled by the neutrino from production to the detector. 

Most measurements are from disappearance experiments. The SNO results may be considered as coming from indirect appearance experiments. At higher energies the OPERA experiment is a direct appearance experiment, while other experiments may obtain indirect appearance signals.

\section{Atmospheric $\nu$ experiments}

After the early indications of an anomaly in atmospheric $\nu$'s \cite{1, 2, 6}, in 1998 Soudan 2, MACRO and SK provided strong indications in favour of $\nu_{\mu} \leftrightarrow \nu_{\tau}$ oscillations. Confirming results were presented by long baseline experiments.

\begin{figure}[ht]
\centering
 {\centering\resizebox*{5.5cm}{!}{\includegraphics{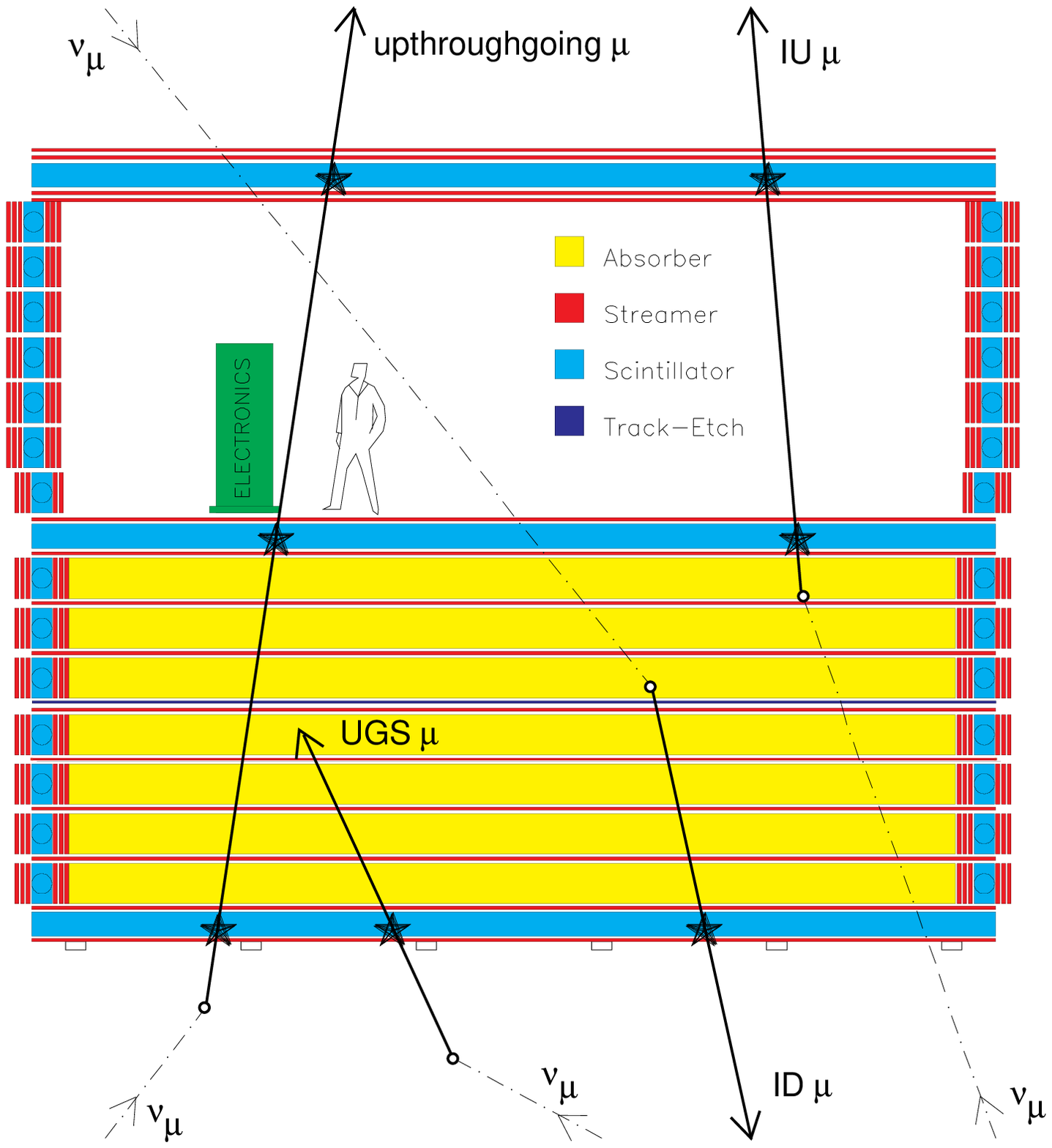}}}
  \hspace{0.5cm}
 {\centering\resizebox*{8.5cm}{!}{\includegraphics{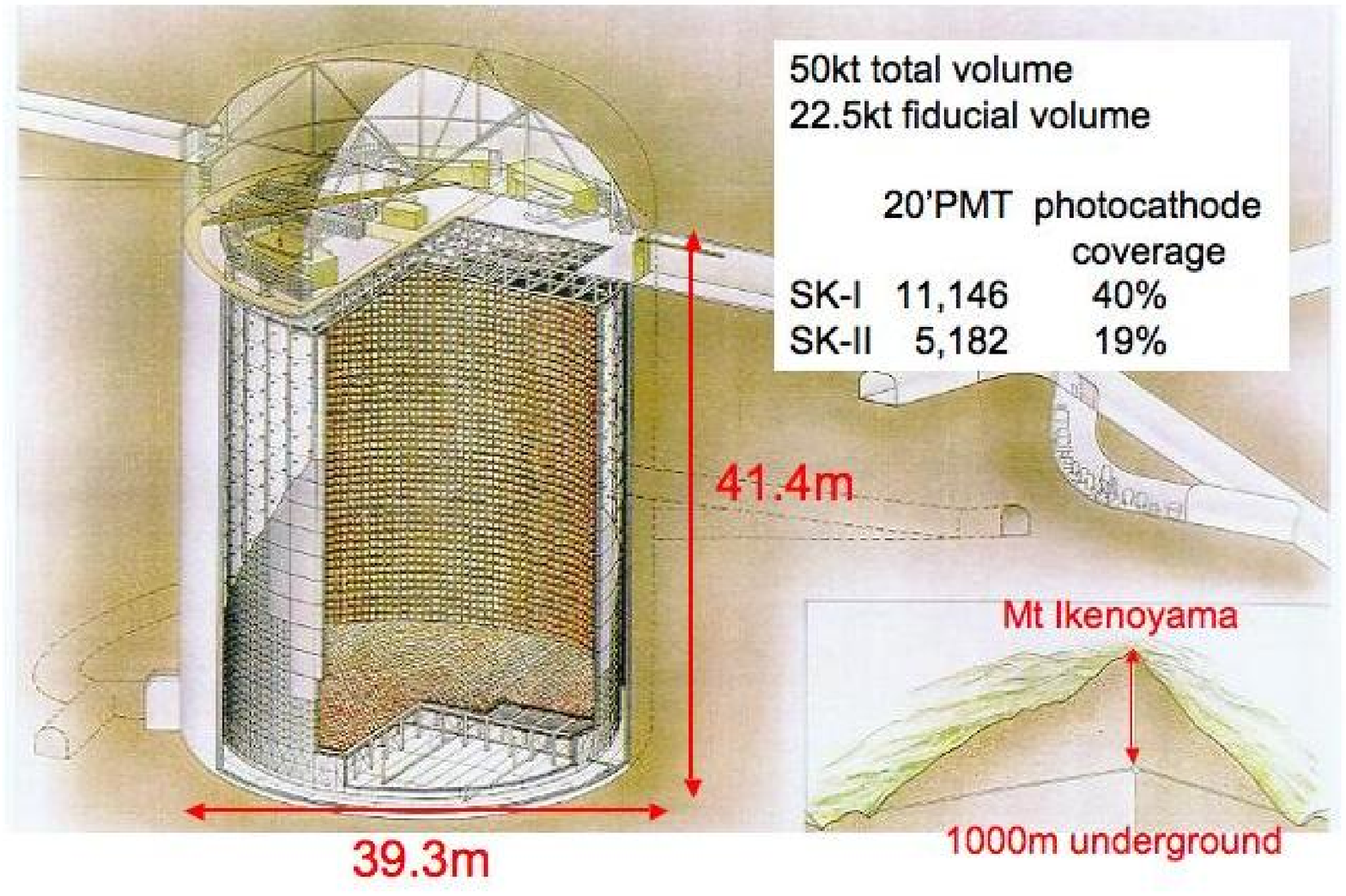}}}
\caption{\small Left) Cross section of the MACRO detector and sketch of event topologies. Right) Schematic layout of the SuperKamiokande detector.}
\label{}
\vspace{-0.5cm}
 \end{figure}

{\bf Soudan 2} used a modular fine grained tracking and showering calorimeter of 963 t located underground in the Soudan Gold mine in Minnesota. The detector was made of 1m $\times$ 1m $\times$ 2.5m modules, surrounded by an anticoincidence shield. The bulk of the mass was 1.6 mm thick corrugated steel sheets interleaved with drift tubes. The Soudan 2 double ratio for the zenith angle range -1 $\leq$ $cos\Theta$ $\leq$ 1 was R' = (N$_\nu$/N$_{e}$)$_{DATA}$/(N$_{\mu}$/N$_{e}$)$_{MC}$ = 0.68 $\pm$ 0.11stat, consistent with $\nu_{\mu}$ oscillations \cite{8}.

{\bf MACRO} detected upgoing $\nu_{\mu}$'s via CC interactions $\nu_{\mu} \rightarrow \mu$; upgoing muons were identified with streamer tubes (for tracking) and scintillators (for time-of-flight measurements), see Fig. 1 left \cite{9, 13}. Events were classified as: {\it Upthroughgoing muons} from interactions in the rock below the detector of $\nu_{\mu}$ with $\langle E \rangle \sim$ 50 GeV. Data were compared with the predictions of the Bartol96 \cite{11}, FLUKA, HKKM01 \cite{12} MCs, Fig. 2 left. The shapes of the angular distributions and the absolute values favoured $\nu_{\mu}$ oscillations with $\Delta$m$^{2}_{23}$ = 0.0025 eV$^{2}$ and maximal mixing. {\it Low energy events. Semicontained upgoing muons} (IU) come from $\nu_{\mu}$ interactions inside the lower apparatus. {\it Up stopping muons} (UGS) are due to external $\nu_{\mu}$ interactions yielding upgoing muons stopping in the detector; {\it the semicontained downgoing muons} (ID) are due to downgoing $\nu_{\mu}$'s interacting in the lower detector \cite{12}. An equal number of UGS and ID events is expected. The average parent neutrino energy is 2-3 GeV. The data are compared with MC predictions without oscillations and show a uniform deficit over the whole angular distribution with respect to the Bartol96 MC predictions.

\begin{figure}[ht]
\centering
 {\centering\resizebox*{7cm}{!}{\includegraphics{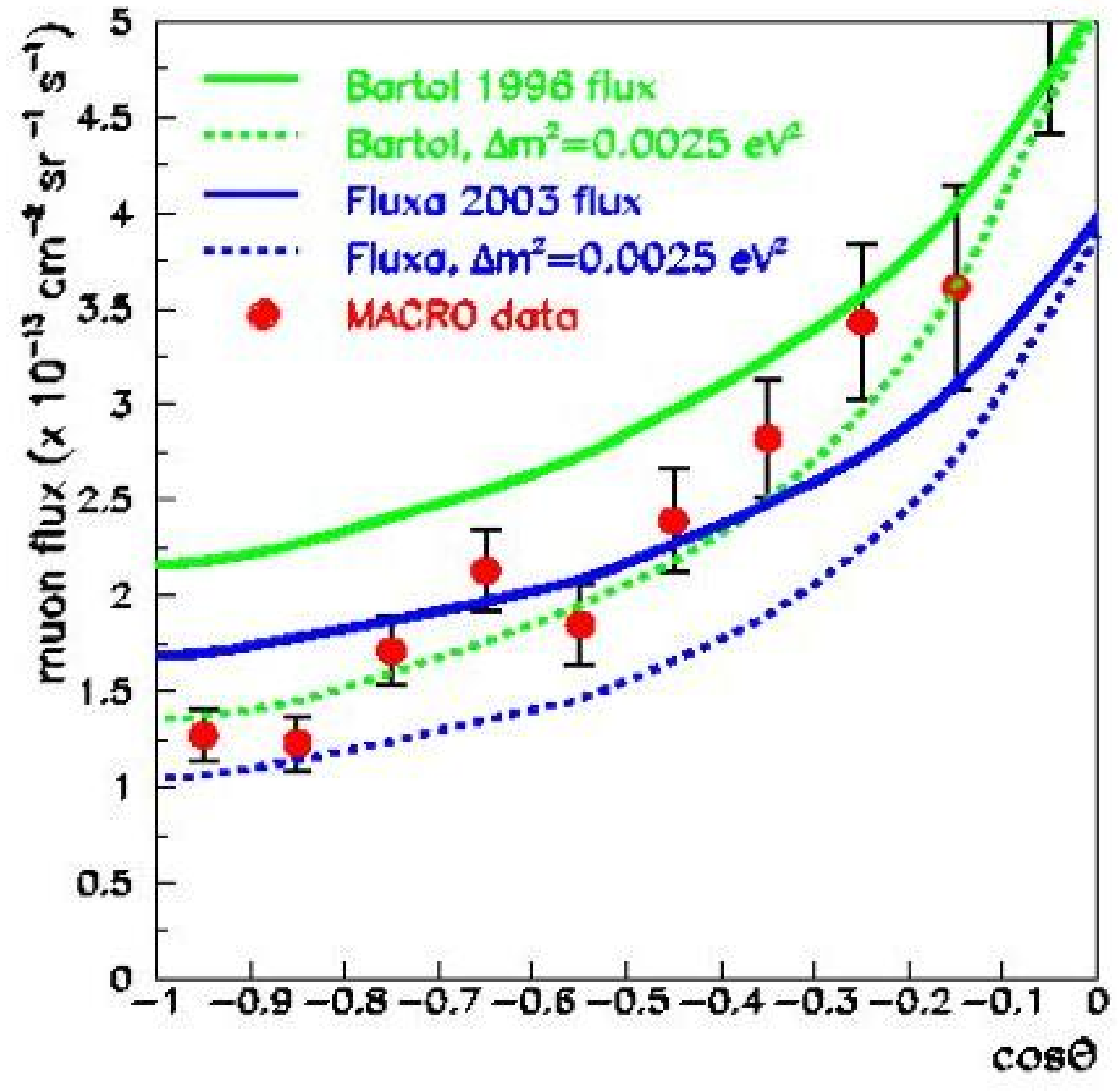}}}
 \hspace{-0.5cm}
 {\centering\resizebox*{7.3cm}{!}{\includegraphics{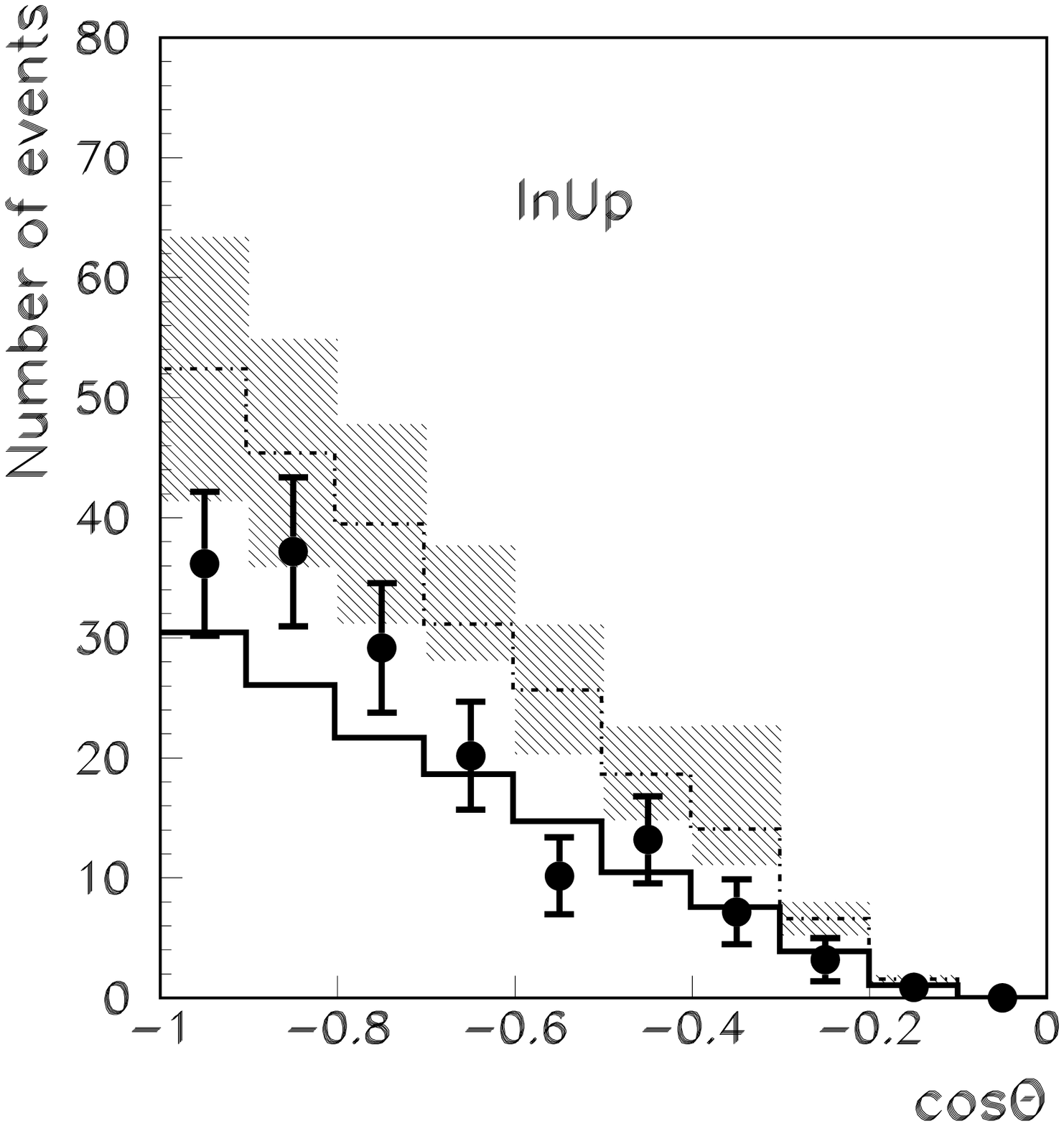}}}
\caption{\small Left) MACRO upthroughgoing muons compared with the oscillated MC predictions of Bartol96 (solid curve), HKKM01 (dash-dotted line), FLUKA fitted to the new CR fit (dashed curve) and FLUKA with the old CR fit (dotted curve). Right) ID events (black points) compared with the non oscillated Bartol96 MC (dashed line) and $\nu_{\mu} \leftrightarrow \nu_{\tau}$ predictions.}
\label{}
 \end{figure}

{\it $\nu_{\mu} \leftrightarrow \nu_{\tau}$ vs $\nu_{\mu} \leftrightarrow \nu_{sterile}$}. Matter effects for $\nu_{\mu}$ with respect to sterile neutrinos, yield different total number and different zenith distributions of upthroughgoing muons. The measured ratio between the events with -1 $<cos\Theta <$ 0.7 and with -0.4$<cos\theta<$0 was R$_{meas}$= 1.38 compared to R$_{\tau}$=1.61 and R$_{sterile}$=2.03. Thus the $\nu_{\mu} \leftrightarrow \nu_{sterile}$ oscillations are exluded at 99.8$\%$ C.L. \cite{13}

{\it $\nu_{\mu}$ energy estimate by Multiple Coulomb Scattering (MCS) of upthroughgoing muons.} The estimate was made through their MCS in the absorbers. The ratios Data/MC$_{no osc}$ as a function of (L/E$_{\nu}$) are in agreement with $\nu_{\mu} \leftrightarrow \nu_{\tau}$ oscillations \cite{13}.

In 2004 in order to reduce the effects of systematic uncertainties in the MCs, MACRO used the following three independent ratios (it was checked that all MCs yield the same predictions):\\
(i) High Energy Data: zenith distribution ratio: R$_{1}$ = N$_{vert}$/N$_{hor}$; (ii) High Energy Data: $\nu_{\mu}$ energy measurement ratio: R$_{2}$ = N$_{low}$ /N$_{high}$; (iii) Low Energy Data: R$_{3}$ = (Data/MC)$_{IU}$/(Data/MC)$_{ID+UGS}$.\\
Fitting the 3 ratios to the $\nu_{\mu} \leftrightarrow \nu_{\tau}$ oscillation formulae, yielded sin$^{2}$2$\theta_{23}$= 1, $\Delta$m$^{2}_{23}$= 2.3 $\cdot$10$^{-3}$ eV$^{2}$. Using Bartol96, one adds the information on absolute fluxes:\\
(iv) High energy data (systematic error $\simeq$17$\%$): R$_{4}$ = N$_{meas}$/N$_{MC}$.\\
(v) Low energy semicontained muons (scale error 21$\%$): R$_{5}$ = N$_{meas}$/N$_{MC}$.\\
The last 2 ratios leave the best fit values unchanged and improve the significance of the oscillations \cite{13}.

\begin{figure}[ht]
\centering
 {\centering\resizebox*{9.1cm}{!}{\includegraphics{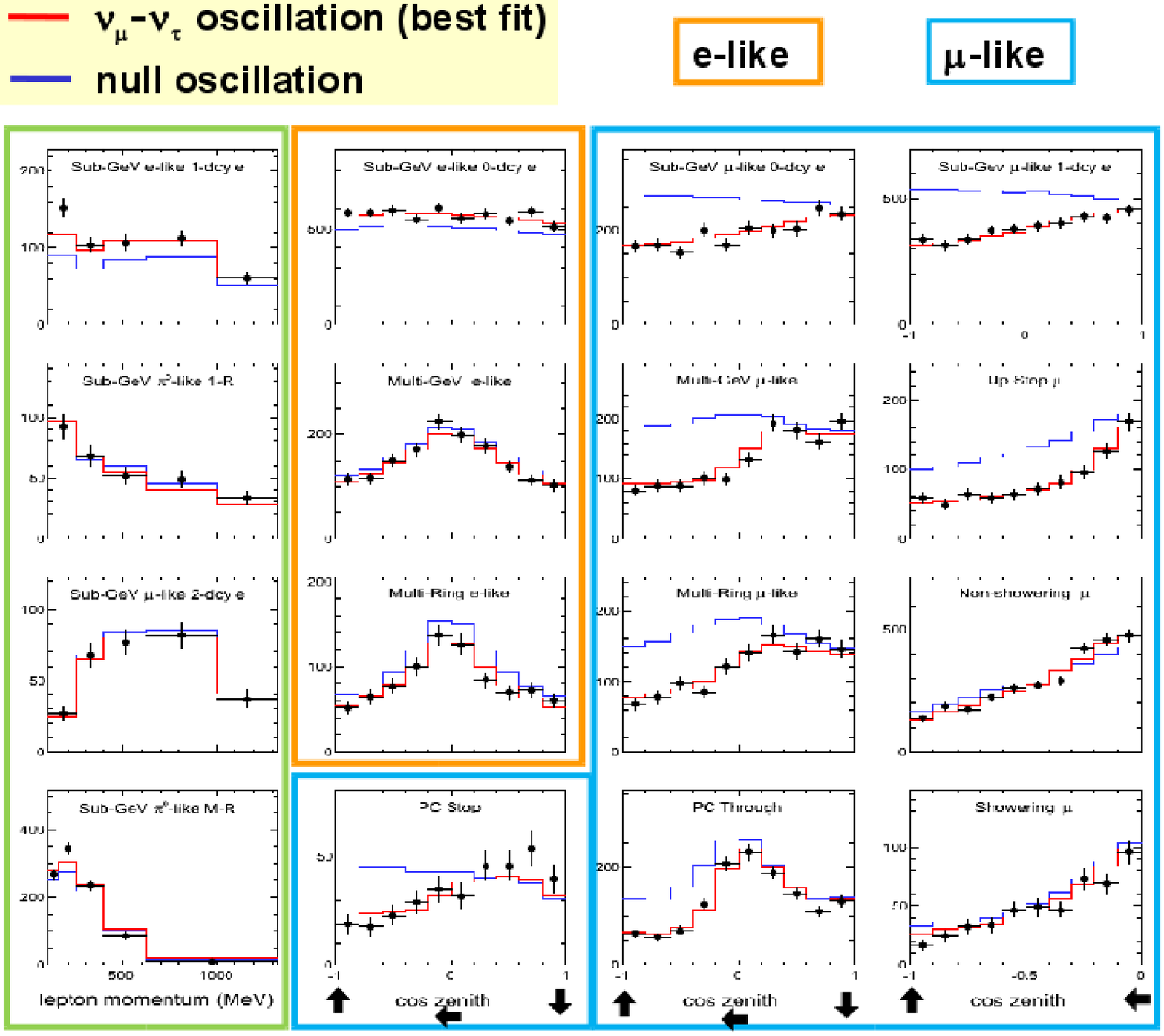}}} 
\hspace{-0.4cm}
 {\centering\resizebox*{7cm}{!}{\includegraphics{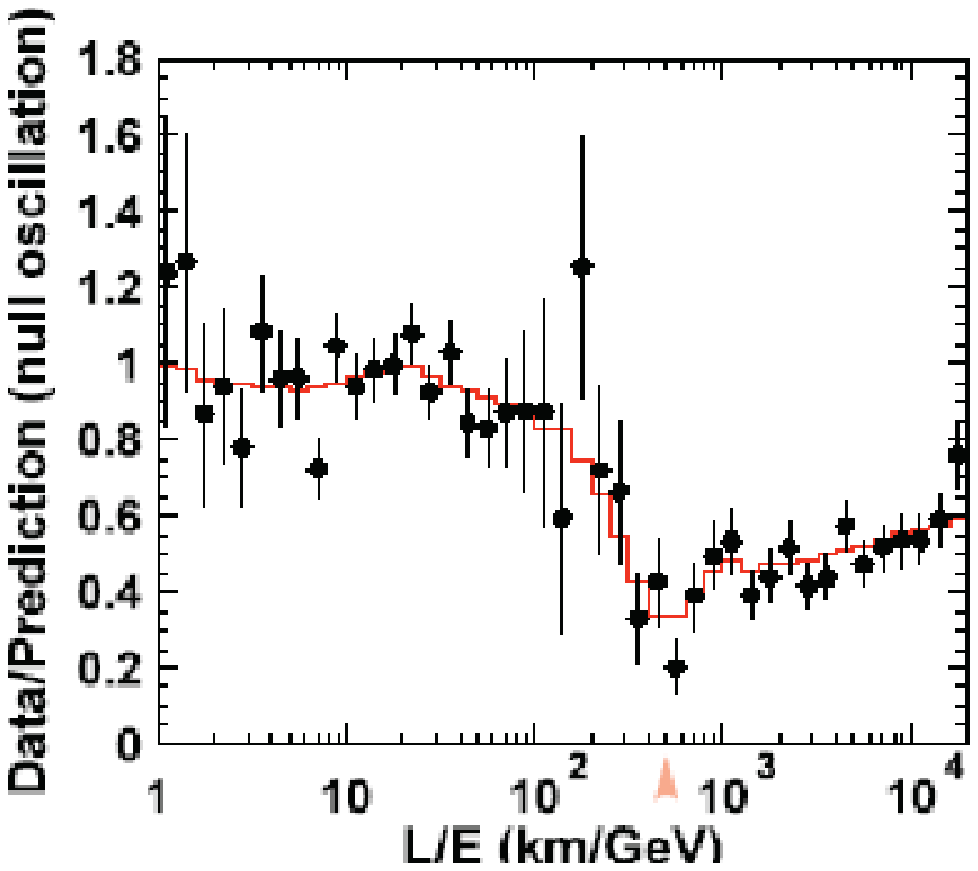}}}
\caption{\small Left) SK zenith distributions (black points) for $e$-like and $\mu$-like sub-GeV, multi-GeV, throughgoing and stopping muons. Solid lines are no oscillation MC predictions. Right) L/E$_{\nu}$ distribution for $\mu$-like events.}
\label{}
 \end{figure}

{\bf SuperKamiokande} (SK) is a large cylindrical water Cherenkov detector of 39 m diameter and 41 m height containing 50 kt of water (fiducial mass 22.5 kt); it is seen by 50-cm-diameter inner-facing phototubes (PMTs), Fig. 1 right. The 2 m thick outer layer of water acts as an anticoincidence; it is seen by smaller outward facing PMTs. The detector is located in the Kamioka mine, Japan. Atmospheric neutrinos are detected in SK by measuring the Cherenkov light generated by the charged particles produced in the neutrino CC interactions with water nuclei. Thanks to the high PMT coverage, the experiment detects events of energies as low as $\sim$5 MeV.

The large detector mass allows to collect a high statistics sample of {\it fully contained} events {\it(FC)} up to $\sim$5 GeV. The FC events yield rings of Cherenkov light on the PMTs. FC events are subdivided into {\it sub-GeV} and {\it multi-GeV} events, with energies below and above 1.33 GeV. FC events include only single-ring events, while {\it multi-ring} ones (MRING) are treated as a separate category. The {\it partially contained} events (PC) are CC interactions with vertex within the fiducial volume and at least a charged particle, typically the $\mu$, exits the detector (the light pattern is a filled circle). {\it Upward-going muons} (UPMU), produced by $\nu_{\mu}$ interacting in the rock before the detector, are subdivided into {\it stopping} ($\langle E \rangle \sim$7 GeV) and {\it throughgoing muons} ($\langle E \rangle \sim$ 70 $\div$ 80 GeV).

Particle identification is performed using likelihood functions to parametrize the sharpness of the Cherenkov rings, which are more diffused for electrons than for muons. The zenith angle distributions for $e$-like and $\mu$-like {\it sub-GeV} and {\it multi-GeV} events are shown in Fig. 3 left. A MC problem exists also in SK: to reduce these problems the normalization is now left as a free parameter.

The ratios $e$-like events/MC do not depend from L/E$_{\nu}$ while $\mu$-like events/MC show a dependence on L/E$_{\nu}$ consistent with the oscillation hypothesis, Fig. 3 right. In 1998 the overall best fit of SK data yielded maximal mixing and $\Delta$m$^{2}_{23}$= 2.3 $\cdot$ 10$^{3}$ eV$^{2}$ (2.5 in 2004). In 2010 SK started fits with three neutrino types: they obtained maximal mixing for $\theta_{23}$ and $\Delta$m$^{2}_{23}$ $\simeq$2.2-2.3 $\cdot$ 10$^{3}$ eV$^{2}$ \cite{14}.

{\bf Exotic oscillations.} MACRO and SuperK data were used to search for sub-dominant oscillations due to Lorentz invariance violation (or violation of the equivalence principle). In the first case there could be mixing between flavor and velocity eigenstates. 90$\%$ C.L. limits were placed in the Lorentz invariance violation parameters $\mid \Delta v \mid$ $<$6 $\cdot$ 10$^{-24}$ at sin$^{2}$2$\theta_{v}$=0 and $\mid \Delta v \mid$ $<$4 $\cdot$ 10$^{-26}$ at sin$^{2}$2$\theta_{v}$=$\pm$1 \cite{15, 16, 17}. 

{\bf Neutrino decay} could be another exotic explanation for neutrino disappearence; SK data do not favour $\nu$ decay and no radiative decay was observed in specific experiments \cite{18}.

New limits were obtained on proton decay and on GUT magnetic monopoles by SK and MACRO.

\section{Long base line $\nu$ experiments}

Atmospheric neutrino oscillations have opened a new window into phenomena beyond the Standard Model of particle physics. Long baseline neutrino experiments allow further insight into $\nu$ physics. The first long baseline $\nu$ beam was the KEK to Kamioka beam (250 km), the 2$^{nd}$ was the Fermilab to the Soudan mine beam (NuMi) (735 km) and the third was the CERN to Gran Sasso (CNGS) (730 km).

The {\bf K2K} experiment, which used SuperKamiokande as the far detector and the low energy $\nu_{\mu}$ beam from KEK to Kamioka, confirmed the results of atmospheric $\nu_{\mu}$ oscillations \cite{19}.

The {\bf MINOS} experiment on the NuMi low energy neutrino beam is a large magnetised steel scintillator tracking calorimeter, complemented by a similar near detector and a calibration detector, Fig 4 left. The experiment obtained important results which confirmed the atmospheric $\nu$ oscillation picture with maximal mixing and $\Delta$m$^{2}_{23}$= 2.35 eV$^{2}$, Fig. 4 right \cite{20}. Recently MINOS reported a possible 40$\%$ difference between $\nu_{\mu}$ and $\overline{\nu_{\mu}}$ oscillation parameters, Fig 4 right \cite{21}. They plan to check this point with new runs, in particular with the $\overline{\nu_{\mu}}$ beam. The MINOS experiment obtained also several interesting limits on sterile neutrinos, on Lorentz invariance and measured the neutrino velocity and the ratio $\mu^{+}$/$\mu^{-}$ for atmospheric muons \cite{22}. The same ratio was also measured by the OPERA experiment \cite{23}, Fig 7.

The {\bf CNGS $\nu_{\mu}$ beam.} \cite{24}. A 400 GeV beam is extracted from the SPS and is transported to the target. Secondary pions and kaons are focused into a parallel beam by 2 magnetic lenses, called horn and reflector. Pions and kaons decay into $\nu_{\mu}$ and $\mu$ in a long decay pipe. The $\mu$'s are monitored in 2 detectors. The mean $\nu_{\mu}$ energy is 17 GeV, the $\overline{\nu_{\mu}}$ contamination $\sim$2$\%$, the $\nu_{e}$ ($\overline{\nu_{e}}$) $<$1$\%$ and the number of $\nu_{\tau}$ is negligible. The muon beam size at the 2$^{nd}$ muon detector at CERN has $\sigma \sim$1 m; this gives a $\nu_{\mu}$ beam size at GS of $\sigma \sim$1 km. The first low intensity test beam was sent to GS in 2006 and 3 detectors (OPERA, LVD and Borexino) obtained their first events. The shared SPS beam sent a pulse of 2 neutrino bursts, each of 10.5 $\mu$sec duration, separated by 50 ms, every 12 s. The $\nu_{\mu}$ beam operated regularly in 2008, 2009 and 2010.

\begin{figure}[ht]
\centering
 {\centering\resizebox*{7.5cm}{!}{\includegraphics{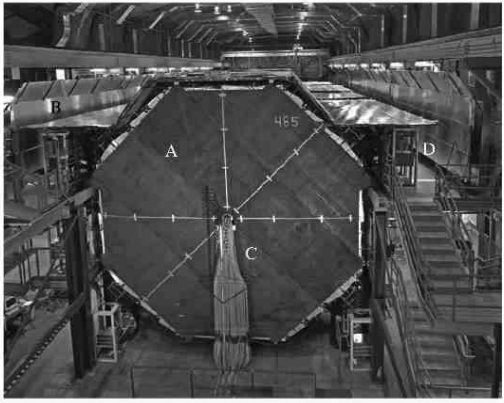}}}
  \hspace{0.5cm}
 {\centering\resizebox*{7.5cm}{!}{\includegraphics{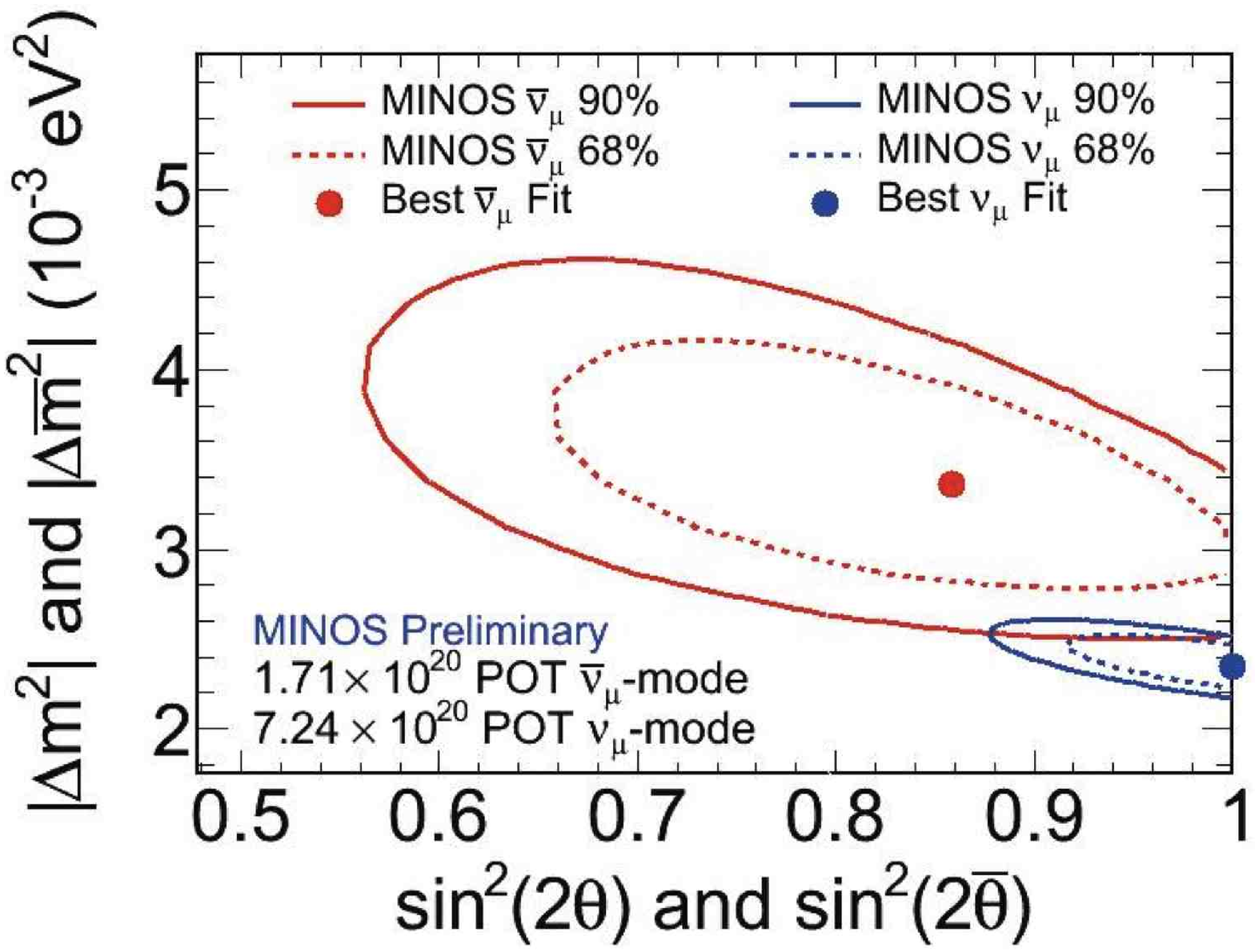}}}
\caption{\small Left: photograph of the MINOS far detector: A indicates the magnetized iron sheets, C is the electric coil, B is an anti coincidence scintillator system. Right: MINOS determination of $\Delta$m$^{2}_{23}$ and sin$^{2}$2$\theta_{23}$ for $\nu_{\mu}$ and $\overline{\nu_{\mu}}$ oscillations.}
\label{}
\vspace{-0.5cm}
 \end{figure}

{\bf LVD} is an array of liquid scintillators with a total mass of 1000 t, designed to search and study $\overline{\nu_{e}}$'s from gravitational stellar collapses. It monitors also the CNGS neutrino beam \cite{25}.

The {\bf ICARUS T600 detector} is an innovative TPC liquid argon detector installed in Hall B of Gran Sasso \cite{24}. It represents an intermediate technical step toward a more massive detector and it can offer several interesting possibilities. Now it is studying neutrino interactions in the CNGS beam with a fiducial mass of 480 t \cite{26}.

{\bf OPERA}, in Hall C at Gran Sasso, is a hybrid emulsion-electronic detector, designed to search for $\nu_{\mu}\rightarrow\nu_{\tau}$ oscillations in appearance mode \cite{27}. The $\nu_{\tau}$ appearance is made by direct detection of the $\tau$ lepton from $\nu_{\tau}$ CC interactions and the $\tau$ lepton decay products. To observe the decays, a spatial resolution of 1 ${\mu}$m is necessary; this is obtained in emulsion sheets interleaved with thin lead target plates (Emulsion Cloud Chamber (ECC)). The OPERA detector, Figure 5, is made of two identical supermodules, each consisting of a target section with 31 target planes of lead/emulsion modules ($bricks$), of a scintillator tracker detector and a muon spectrometer.

\begin{figure}[ht]
\centering
 {\centering\resizebox*{10cm}{!}{\includegraphics{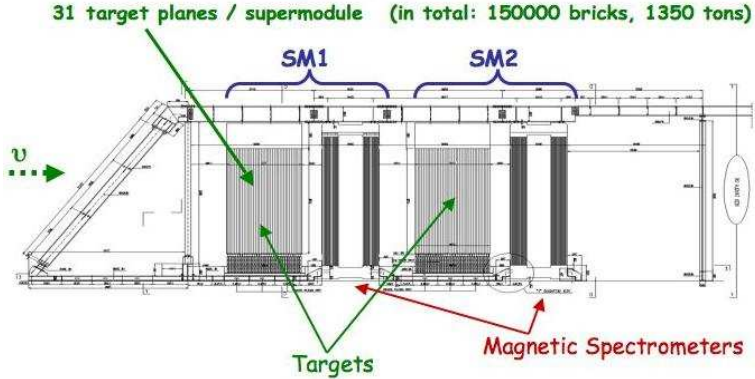}}}
\caption{\small Schematic of the OPERA experiment at Gran Sasso}
\label{fig:fig16}
 \end{figure}

\begin{figure}[h]
\centering
 {\centering\resizebox*{6.2cm}{!}{\includegraphics{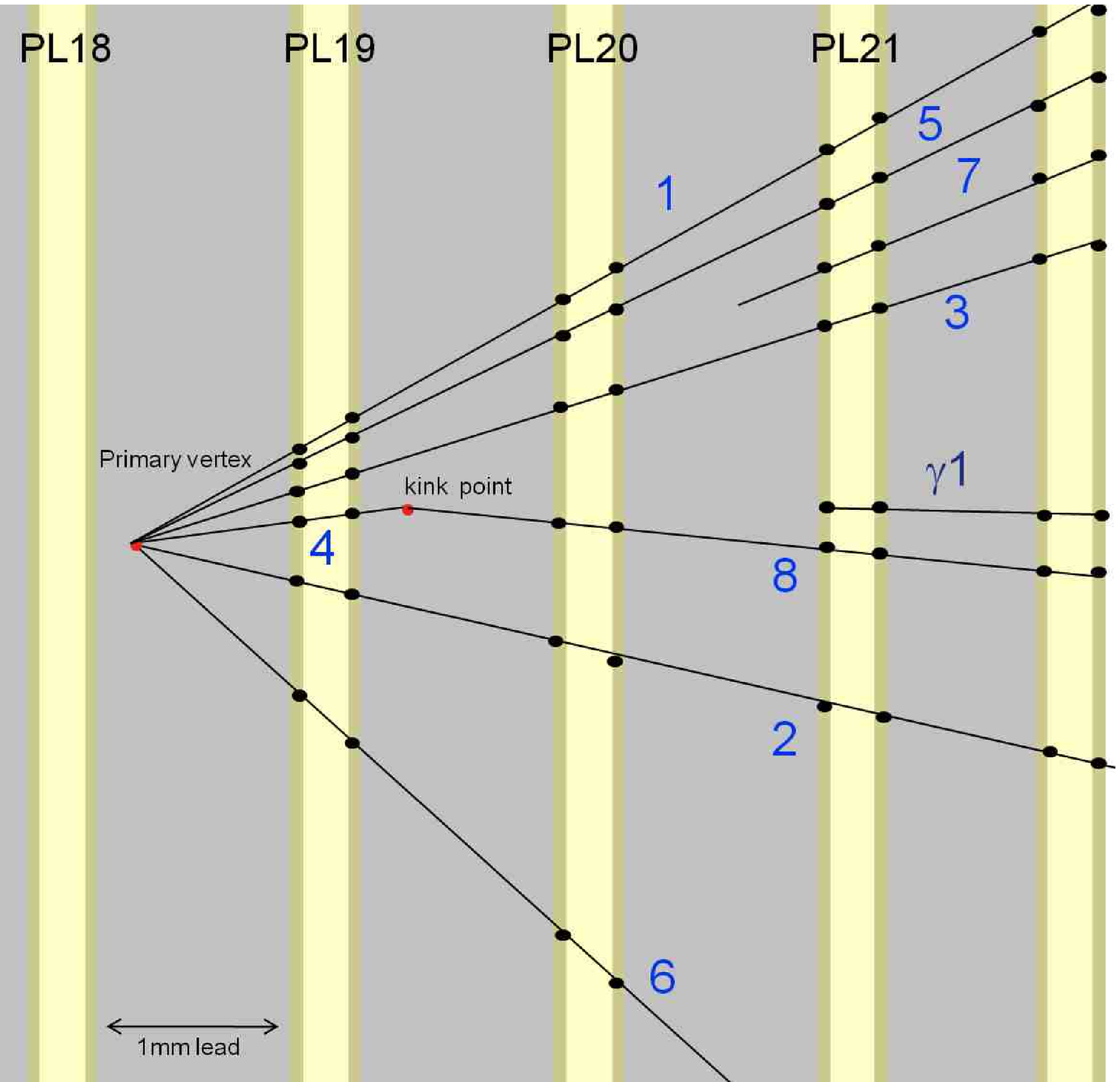}}}
 {\centering\resizebox*{7.1cm}{!}{\includegraphics{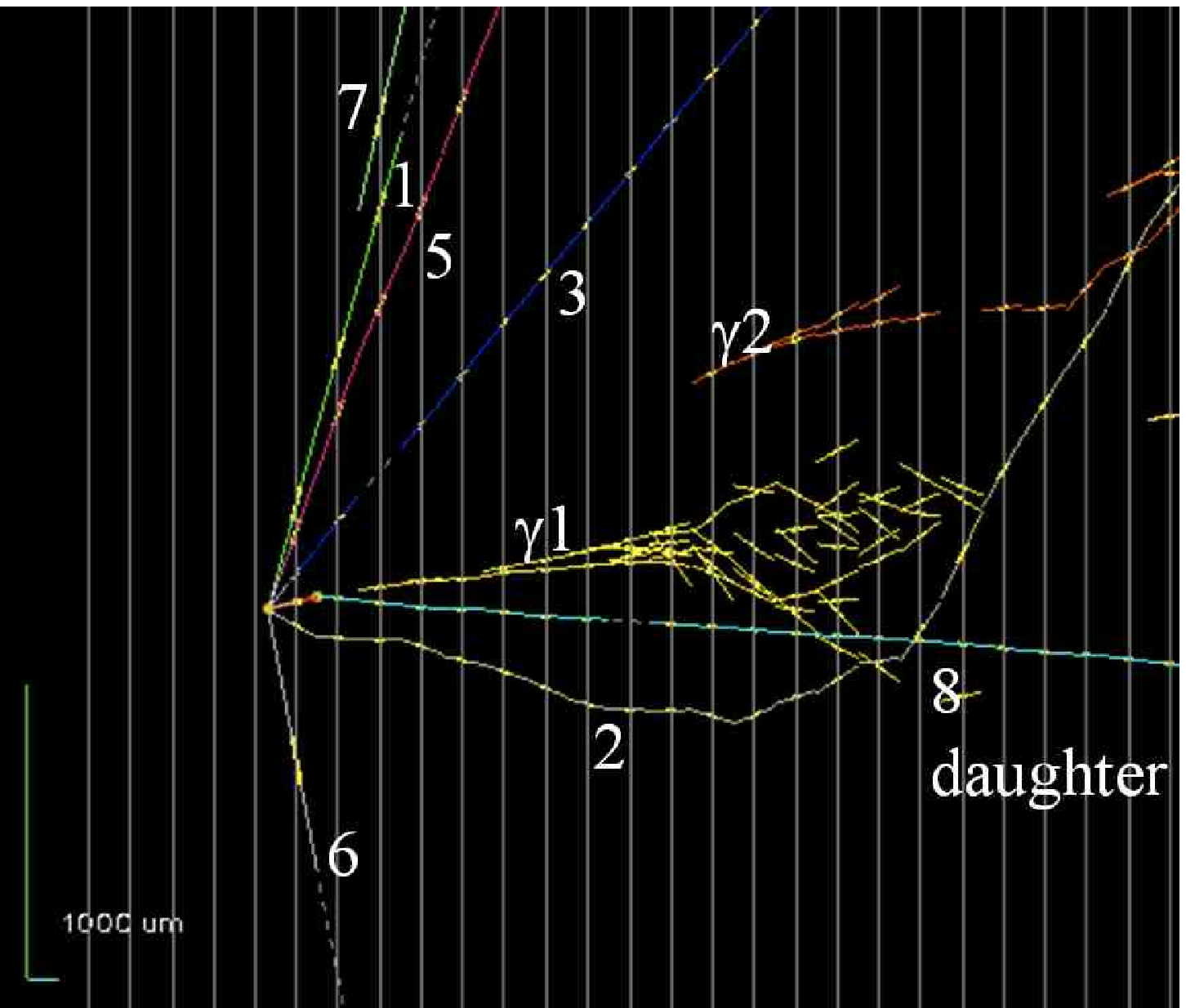}}}
\caption{\small Display of the first $\nu_{\mu}\rightarrow\nu_{\tau}$ candidate event from the OPERA experiment.}
\label{fig:fig18a}
 \end{figure}

The muon spectrometer consists of 2 iron magnets instrumented with Resistive Plate Chambers (RPC) and precision drift tubes. Each magnet is an 8 $\times$ 8 m$^{2}$ dipole with a field of 1.52 T in the upward direction on one side and in the downward direction on the other side. Precision drift tubes, measure the muon track coordinates in the horizontal plane. The muon spectrometer has a $\Delta$p/p$\sim$0.25 for muon momenta of $\sim$25 GeV/c. The basic target module is a $brick$ consisting of 56 lead plates (1 mm thick) and 57 emulsion layers. A brick has a size of 10.2 $\times$ 12.7 cm$^{2}$, a depth of 7.5 cm and a weight of 8.3 kg. Two additional emulsion sheets, the changeable sheets (CS), are glued on its downstream face. Within a brick, the achieved space resolution is $<$1 $\mu$m and the angular resolution is $\sim$2 mrad. Walls of Target Tracker Scintillators provide the $\nu$ interaction trigger and identify the brick in which the interaction took place. The bricks were made by the Brick Assembly Machine (BAM) and are handled by the Brick Manipulator System (BMS). A fast automated scanning system with a scanning speed of $\sim$20 cm$^{2}$/h per emulsion (each 44 $\mu$m thick) is needed to cope with the analyses of so many emulsions. For this purpose were developed the European Scanning System \cite{28} and the Japanese S UTS \cite{29}. An emulsion is placed on a holder and the flatness is guaranteed by a vacuum system. By adjusting the focal plane of the objective, 16 tomographic images of each field of view are taken at equally spaced depths. The images are digitized, sent to a vision processor, analyzed to recognize sequences of aligned grains. The 3-dimensional structure of a track in an emulsion layer ({\it microtrack}) is reconstructed by combining clusters belonging to images at different levels. Each microtrack pair is connected across a plastic base to form the {\it basetrack}. A set of base tracks forms a {\it volume track}. 

\begin{figure}[ht]
\centering
 {\centering\resizebox*{9.8cm}{!}{\includegraphics{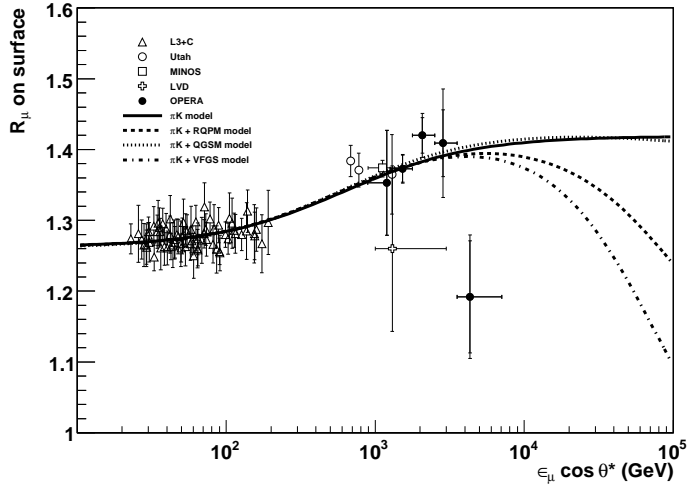}}}
\caption{\small Compilation of ratios $R_{\mu}=N_{\mu^{+}}/N_{\mu^{-}}$ in cosmic rays measured at ground level and underground. Notice in particular the data obtained at large $\epsilon_{\nu}\cos\theta^{\star}$ by MINOS, LVD and OPERA.}
\label{fig:fig18bis}
 \end{figure}

Recently OPERA found a first candidate event $\nu_{\mu}\rightarrow \nu_{\tau}$ \cite{30}. We recall the main phases of the measurements. The electronic detector found one event which had the features of a NC interaction, that is without a muon. The Target Tracker scintillators individuated the brick in which the interaction occurred. The CS of the brick were removed and developed in the underground lab. Their analysis corroborated the TT findings and gave the precise location of some tracks. The brick was then brought to the outside lab, exposed to cosmic ray muons (in order to have more precise emulsion alignments) and developed. The emulsion sheets were later shipped to one of the outside laboratories of the collaboration. There, measurements were made according to the following scheme: starting from the CS, tracks were reconstructed in the ``scan back'' procedure. The vertex of the interaction was located; the measurements proceeded with the ``total scan'' in one cm$^{2}$ around the vertex, which found several tracks. This was followed by the ``decay search'' and ``scan forth'' procedures to better determine the track with a kink and measure all the tracks of the event. Other bricks were extracted and measured to check for interactions, for black tracks of nuclei and for gamma rays. The event was visually checked and was sent to a second lab for a repetition of the measurements. The averages of the 2 measurements were used for track kinematics analysis, determination of the momentum of each track by multiple Coulomb scattering methods, and kinematic analysis of the candidate event, Fig. 6. The decay of the short track was compatible with $\tau^{-}\rightarrow\rho^{-}\nu_{\tau}$. The background from many sources was estimated. The probability that the background fluctuates to one event is few percent giving a statistical significance of slightly over 2$\sigma$ for the observation of 1 real event. One now waits for further events!\\

\section{Conclusions. Outlook}

The atmospheric neutrino anomaly became in 1998 the atmospheric neutrino oscillation hypothesis, which was later confirmed with more data and from the new data from the first two long baseline neutrino experiments. All the experiments were disappearance experiments. All agree with maximal mixing and with the $\Delta$m$^{2}_{23}$ values indicated below in the 2 neutrino oscillation scenario (for SK also in the 3 neutrino scenario \cite{14}): \\
	- In 1998: 2.3 (SK), 2.5 (MACRO)\\
	- In 2004: 2.5 (SK), 2.3 (MACRO)\\
	- In 2006: 2.8 (K2K)\\
	- In 2008: 2.43$\pm$0.12 (MINOS)\\
	- In 2010: 2.35$^{+0.11}_{-0.08}$ (MINOS),...(2.2-2.3)(SK 3-flavors)(2.51 and 2.11 at the Neutrino 2010 Conference).\\
The variations in time and the differences among the values from the main experiments are small and well within their experimental uncertainties.

Now there is also one event from the OPERA appearance experiment and one is waiting for more such events.   

Solar and atmospheric neutrino oscillations provided the first indication for physics beyond the Standard Model of Particle Physics \cite{31}. One needs further data to clear several points, like the recent results of MINOS and the results from the experiments MiniBooNE \cite{32} and LSND \cite{33}. 

New experiments, like T2K \cite{34} and Nova \cite{35}, should improve considerably the situation, also using new particle production data \cite{36}. The main aim of the new experiments is the determination of the mixing angle $\theta_{13}$ and the search for a possible CP violation in the leptonic sector \cite{37, 38}.

\vspace{0.5cm}

{\bf Acknowledgments.} {I acknowledge many colleagues at CERN, LNGS and Bologna for discussions and advices. I thank drs. M. Errico and R. Giacomelli for technical help. }

\bibliographystyle{plain}

\begin{thebibliography}{99}
\small{
\bibitem{1} R. Becker-Szendy et al., Phys. Rev. D {\bf 46}, 3720 (1992).
\vspace{-0.3cm}
\bibitem{2} Y. Fukuda et al., Phys. Lett. B {\bf 335}, 237 (1994); K.S. Hirata et al., Phys. Lett. B {\bf 205}, 416 (1988).
\vspace{-0.3cm}
\bibitem{3} M. Aglietta et al., 23rd ICRC Proc. {\bf 2014}, 446 (1993); Europhys. Lett. {\bf 8}, 611 (1989).
\vspace{-0.3cm}
\bibitem{4} K. Daum et al., Z. Phys. C {\bf 66}, 417 (1995); C. Berger et al., Phys. Lett. B {\bf 227}, 489 (1889).
\vspace{-0.3cm}
\bibitem{5} S. Mikheyev et al., Phys. Lett. B {\bf 391}, 491 (1997); 5$^{th}$ Workshop, LNGS Gran Sasso, Italy.
\vspace{-0.3cm}
\bibitem{6} S. Ahlen et al., Phys. Lett. B {\bf 357}, 481 (1995); Nucl. Instrum. Meth. A {\bf 324}, 337 (1993).
\vspace{-0.3cm}
\bibitem{7} W. W. M. Allison et al., Phys. Lett. B {\bf 391}, 491 (1997).
\vspace{-0.3cm}
\bibitem{8} E. Peterson for the Soudan2 Coll., Nucl. Phys. B {\bf 77}, 111 (1999); W. W. M. Allison et al., Phys. Lett. B {\bf 449}, 137 (1999).
\vspace{-0.3cm}
\bibitem{9} M. Ambrosio et al., Phys. Lett. B {\bf 434}, 451 (1998); Phys. Lett. B {\bf 478}, 5 (2000); Phys. Lett. B {\bf 517}, 59 (2001). P. Bernardini, hep-ex/9809003 (1998). F. Ronga et al., hep/9810008 (1998). 
\vspace{-0.3cm}
\bibitem{10} Y. Fukuda et al., Phys. Rev. Lett. 81, 1562 (1998); Phys. Lett. B {\bf 433}, 9 (1998); Phys. Rev. Lett. {\bf 85}, 3999 (2000); Phys. Rev. Lett. {\bf 82}, 2644 (1999); Y. Ashie et al., Phys. Rev. Lett. {\bf 93}, 101801 (2004); Phys. Rev.  D {\bf 71}, 11 (2005).
\vspace{-0.3cm}
\bibitem{11} V. Agrawal et al., Phys. Rev. D {\bf 53}, 1314 (1996); M. Honda et al., Phys. Rev. D {\bf 52}, 4985 (1995).
\vspace{-0.3cm}
\bibitem{12} M. Honda et al., Phys. Rev. D {\bf 64}, 053011 (2001); Phys. Rev. D {\bf 70}, 043008 (2004). G. Battistoni et al., Astrop. Phys. {\bf 19}, 269 (2003).
\vspace{-0.3cm}
\bibitem{13} M. Ambrosio et al., Eur. Phys. J. {\bf C36} 323 (2004); hep-ex/0206027; Phys. Lett. {\bf B566} 35 (2003).
\vspace{-0.3cm}
\bibitem{14} Y. Ashie et al., Phys. Rev. Lett. {\bf 93} 101801 (2004); Phys. Rev. Lett. {\bf D71} 112005 (2005). Y. Takeuchi et al., Proc. Neutrino 2010 Conf., Athens.
\vspace{-0.3cm}
\bibitem{15} G. Battistoni et al., Phys. Lett. {\bf B615} 14 (2005).
\vspace{-0.3cm}
\bibitem{16} G. L. Fogli et al., Phys. Rev. {\bf D60} 053006 (1999); Phys. Rev. {\bf D59} 117303 (1999).
\vspace{-0.3cm}
\bibitem{17} P. Lipari and M. Lusignoli, Phys. Rev. {\bf D58} 073005 (1998). Y. Fukuda et al., Phys. Rev. Lett. {\bf 85} 3999 (2000).
\vspace{-0.3cm}
\bibitem{18} S. Cecchini et al.,  Astropart. Phys. {\bf 21} 183 (2004); Astropart. Phys. {\bf 21} 35 (2004); arXiv:0912.5086 [hep-ex].
\vspace{-0.3cm}
\bibitem{19} M. H. Ahn et al., Phys. Rev. Lett. {\bf 90} 041801 (2003); Phys. Rev. {\bf D74} 072003 (2006). E. Aliu et al., Phys. Rev. Lett. {\bf 94} 081802 (2005).
\vspace{-0.3cm}
\bibitem{20} P. Adamson et al., Phys. Rev. Lett. {\bf 101} 131802 (2008); Phys. Rev. {\bf D77} 072002 (2008). D. G. Michael et al., Phys. Rev. Lett. {\bf 97} 191801 (2006).
\vspace{-0.3cm}
\bibitem{21} P. Volhe for the MINOS Coll., Neutrino 2010 Conf., Athens.
\vspace{-0.3cm}
\bibitem{22} P. Adamson et al., Phys. Rev. {\bf D76} 072005 (2007); Phys. Rev. Lett. {\bf 101} 151601 (2008); Phys. Rev. {\bf D76} 072003 (2007).
\vspace{-0.3cm}
\bibitem{23} N. Agafonova et al., Eur. Phys. J. {\bf C67} 25 (2010).
\vspace{-0.3cm}
\bibitem{24} CNGS project: http://proj-cngs.web.cern.ch/proj-ccngs/. G. Giacomelli, arXiv: physics/0703247 [physics.ins-det]. E. Gschwendtner et al., EuCARD-CON-2009-014, 11$^{th}$ ICATPP Conf., Como, Italy.
\vspace{-0.3cm}
\bibitem{25} L. Selvi for the LVD, Collaboration, AIP Conf. Proc. 944, pp 7 (2007), doi 10.1063/1.2818553.
\vspace{-0.3cm}
\bibitem{26} A. Guglielmi for the ICARUS Coll. Neutrino 2010 Conf., Athens.
\vspace{-0.3cm}
\bibitem{27} R. Acquafredda et al., New J. Phys. {\bf 8} 303 (2006); JINST {\bf 4} P04018 (2009); JINST {\bf 4} (2009) P06020.\\ A. Anokhina et al., JINST {\bf 3} P07005 (2008); JINST {\bf 3} P07002 (2008).
\vspace{-0.3cm}
\bibitem{28} L. Arrabito et al., Nucl. Instrum. Meth.  {\bf A568} 578 (2006); 2007 JINST 2 P02001; 2007 JINST 2 P05004. \\M. De Serio et al., Nucl. Instrum. Meth. {\bf A554} 132 (2005).
\vspace{-0.3cm}
\bibitem{29} T. Fukuda et al., JINST {\bf 5} P04009 (2010).
\vspace{-0.3cm}
\bibitem{30} N. Agafonova et al., Phys. Lett. {\bf B691} 138 (2010).
\vspace{-0.3cm}
\bibitem{31} G. Giacomelli, Radiat. Meas. {\bf 44} 826 (2009). 
\vspace{-0.3cm}
\bibitem{32} A. A. Aguilar-Arevalo, atXiv: 1007.1150v1 [hep-ex].
\vspace{-0.3cm}
\bibitem{33} C. Athanossopoulos et al., Phys. Rev. Lett. {\bf 81} 1774 (1998); Phys. Rev. {\bf D64} 112007 (2001).  
\vspace{-0.3cm}
\bibitem{34} T. Kobayashi, (Status of T2K) Proc. Neutrino 2010 Conf., Athens. 
\vspace{-0.3cm}
\bibitem{35} K. Heller, (No$\nu$a experiment status) Proc. Neutrino 2010 Conf., Athens.
\vspace{-0.3cm}
\bibitem{36} M. G. Catanesi et al., Eur. Phys. J. {\bf C54} 37 (2008). A. Blondel, Proceedings of the Neutrino 2010 Conf., Athens.
\vspace{-0.3cm}
\bibitem{37} M. Mezzetto and T. Schwetz, arXiv: 1003.5800v2 [hep-ph] (2010).
\vspace{-0.3cm}
\bibitem{38} G. Giacomelli, (Atmospheric $\nu$ and long baseline $\nu$ experiments), Lecture at the 2007 Carpatian Summer School of Physics, Sinaia, arXiv: 0712.2126 [hep-ex], AIP Conf. Proceed. 972 (2008) 412 pag. 252. 
}

\end{thebibliography}

\end{document}